\documentclass[epjCONF]{svjour}
\usepackage{graphics}
\usepackage[varg]{txfonts} 
\usepackage[latin1]{inputenc}
\session-title{MESON2012 - 12th International Workshop on Meson Production, 
Properties and Interaction}
\begin{document}
\title{Exclusive production of $\pi^+\pi^-$ and $\pi^0\pi^0$ pairs
in photon-photon and in ultrarelativistic heavy ion collisions}
\author{M.~K{\l}usek-Gawenda\inst{1}\fnmsep\thanks{\email{mariola.klusek@ifj.edu.pl}} 
\and A.~Szczurek\inst{1,}\inst{2}}
\institute{Institute of Nuclear Physics PAN, PL-31-342 Cracow, Poland 
\and University of Rzesz\'ow, PL-35-959 Rzesz\'ow, Poland}
\abstract{
The $\gamma\gamma\to\pi\pi$ reactions are discussed. 
To describe those processes, we include dipion continuum, 
resonances, high-energy pion-pion rescatterings, 
$\rho$ meson exchange and pQCD Brodsky-Lepage mechanisms.
The cross section for the production of pion pairs
in photon-photon collisions in peripheral heavy ion collisions
is calculated with the help of Equivalent Photon Approximaption
(EPA) in the impact parameter space. 
We show predictions at $\sqrt{s_{NN}}=3.5$ TeV
which could be measured e.g. by the ALICE collaboration at the LHC.
} 
\maketitle
\section{Introduction}
%
Due to the large charge of colliding nuclei,
ultrarelativistic collisions of heavy ions provide 
an opportunity to study photon-photon collisions. 
We focus on the understanding of 
both $\gamma\gamma\to\pi^+\pi^-$
and $\gamma\gamma\to\pi^0\pi^0$ processes 
in the whole energy range 
of the exprimental data.
Different mechanisms may contribute in general.
Correct form of the elementary cross section
is necessary to calculate the prediction for future 
nucleus-nucleus experiments.  

\section{Elementary cross section}
%
The $\gamma\gamma\to\pi\pi$ processes
are fairly complicated.
So far nobody was able to describe both
the production  of neutral and charged pions
at low and intermediate energies.
We discuss those processes
starting from kinematical threshold 
($W=2m_\pi$) up to about $W=6$ GeV. 
\begin{figure}[h]
\centering
\resizebox{0.85\columnwidth}{!}{%
  \includegraphics{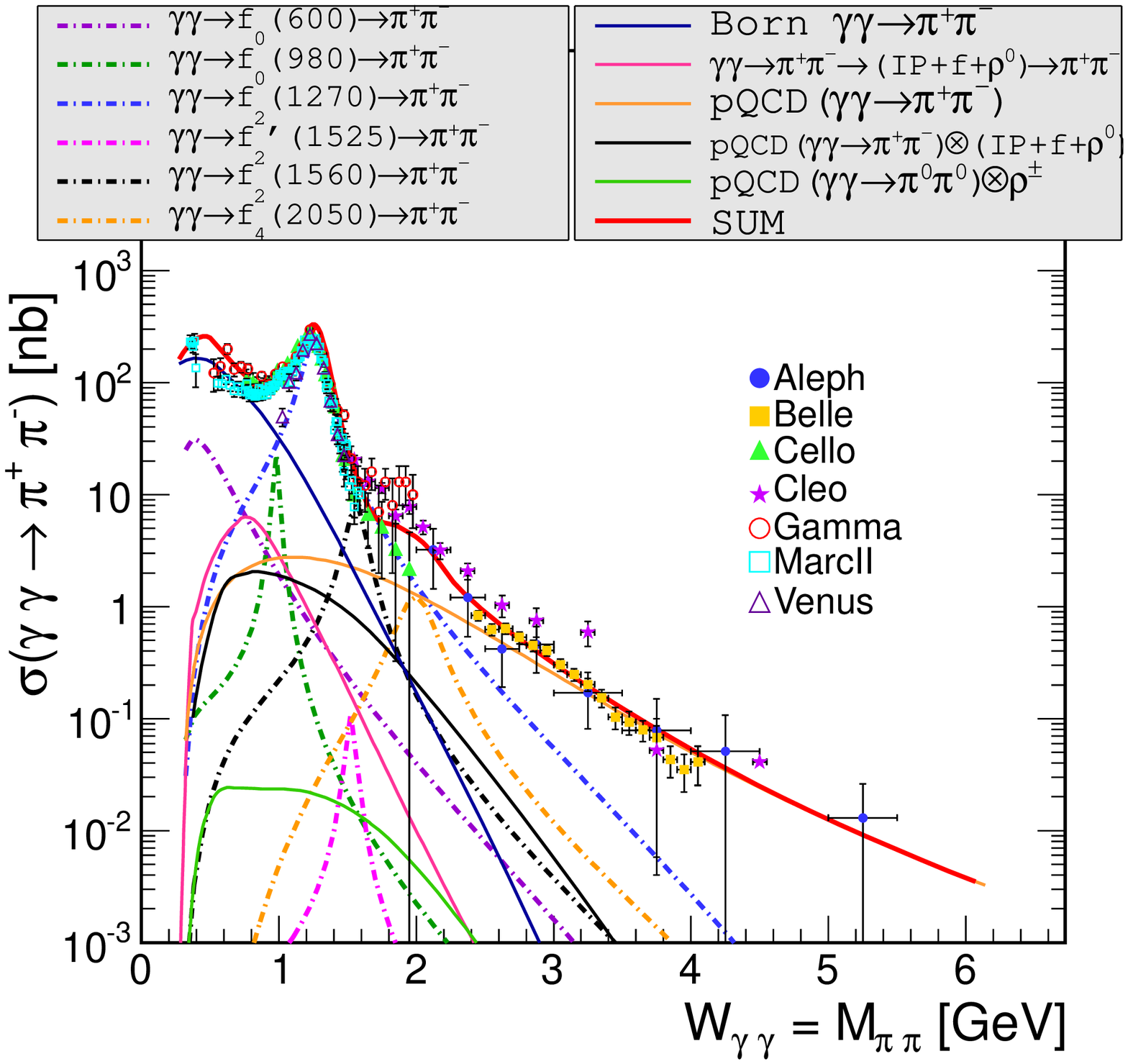}
  \includegraphics{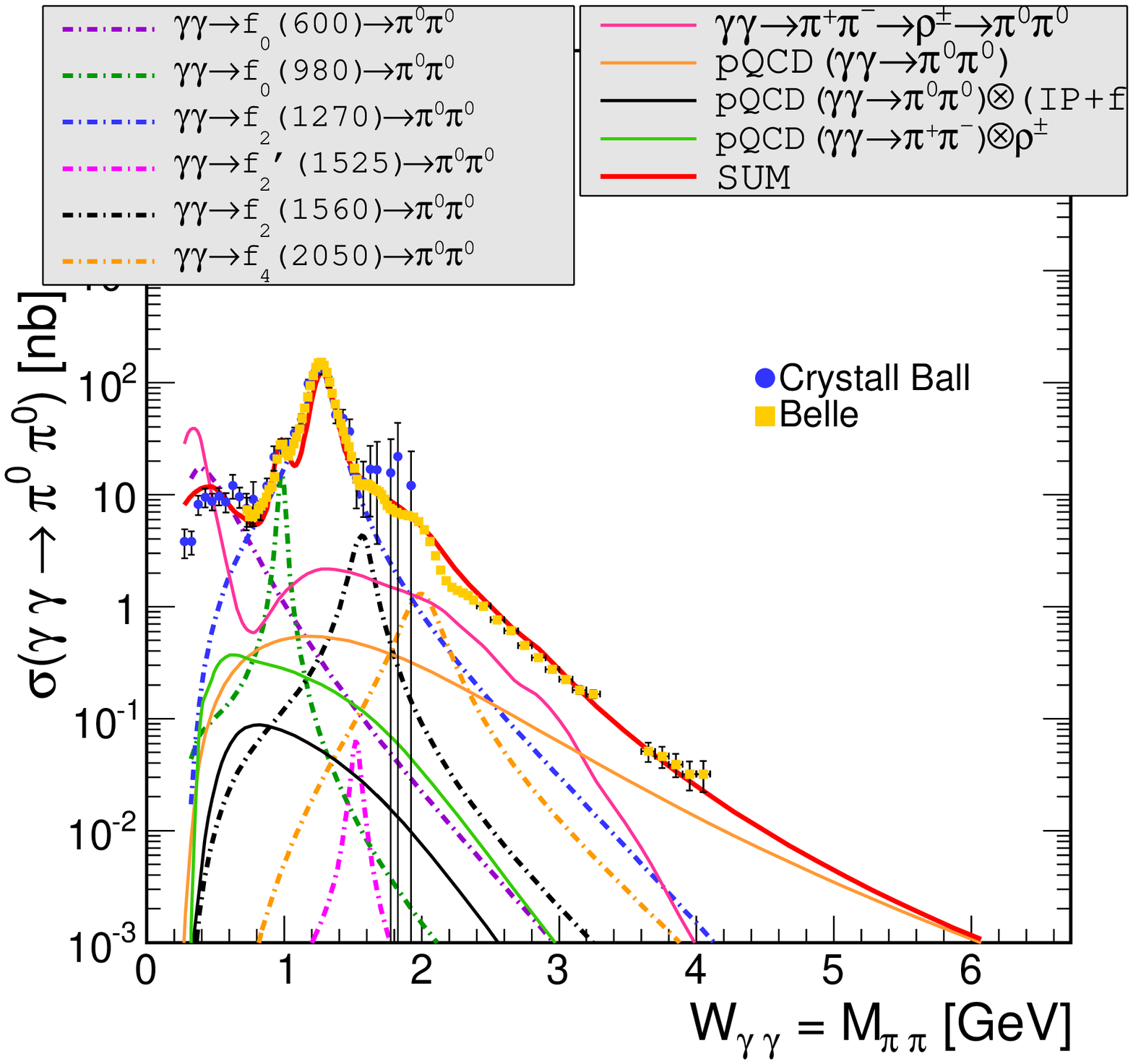} }
\caption{Energy dependence of 
the $\gamma\gamma\to\pi^+\pi^-$ (left panel) and 
the $\gamma\gamma\to\pi^0\pi^0$ (right panel)
cross section. 
Experimental data points have been obtained 
by different groups.}
\label{fig:elementary}       
\end{figure}
\begin{figure}[h]
\centering
\resizebox{0.95\columnwidth}{!}{%
  \includegraphics{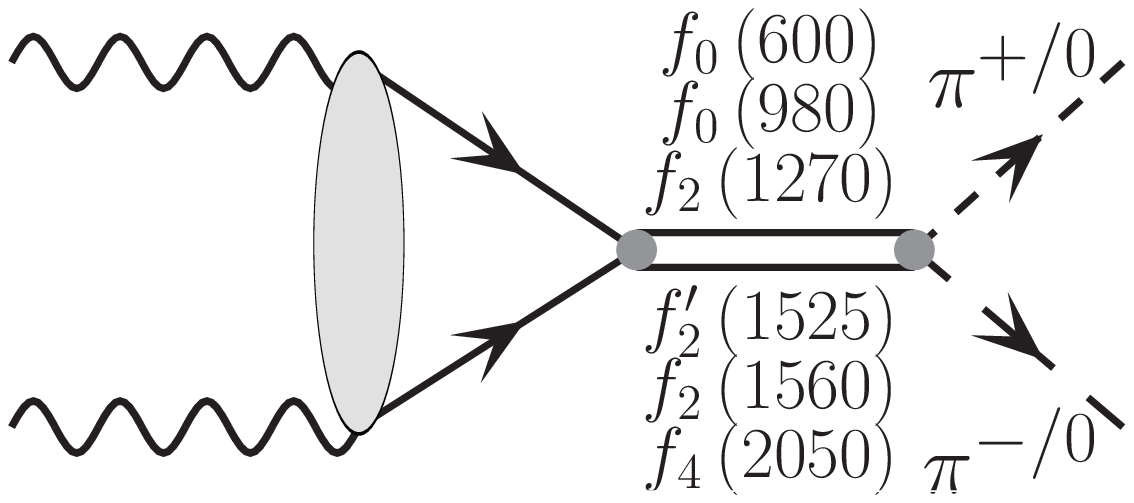}
\hspace{8cm}
  \includegraphics{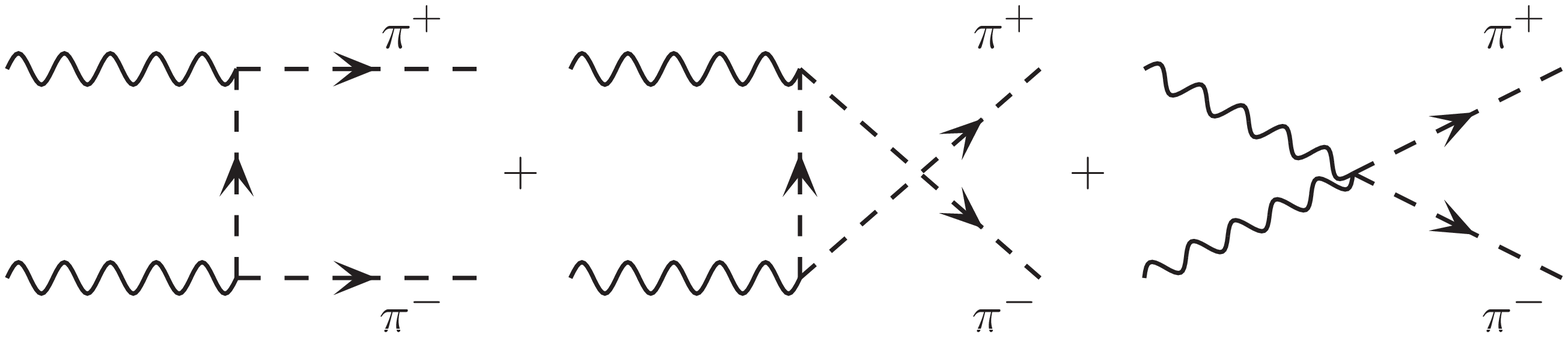} }
\caption{Feynman diagrams describing
the $\gamma\gamma\to\mbox{resonances}\to\pi^{+/0}\pi^{-/0}$ amplitude (left panel)
and the Born amplitude for $\gamma\gamma\to\pi^+\pi^-$ (right panel).}
\label{fig:el1}       
\end{figure}
\begin{figure}[h]
\centering
\resizebox{0.6\columnwidth}{!}{%
  \includegraphics{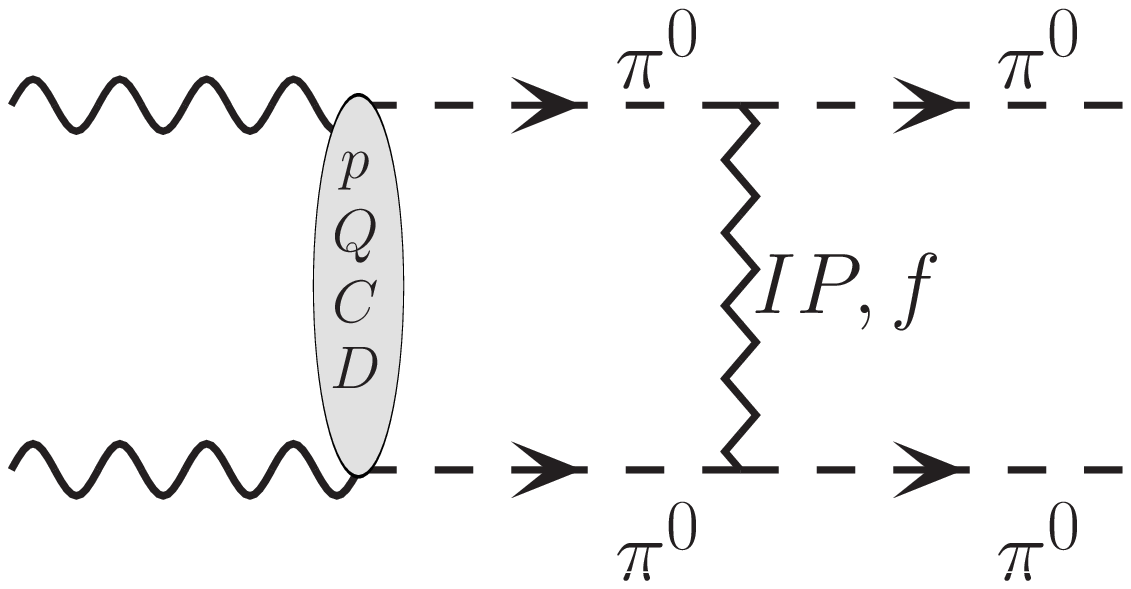}
  \includegraphics{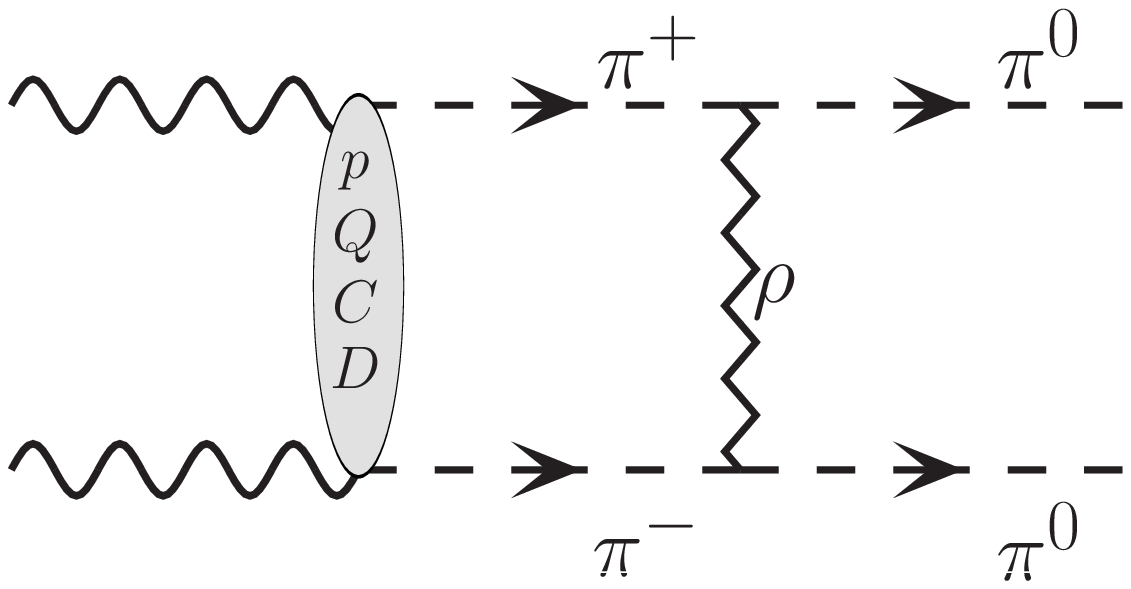}}
\resizebox{0.8\columnwidth}{!}{%
  \includegraphics{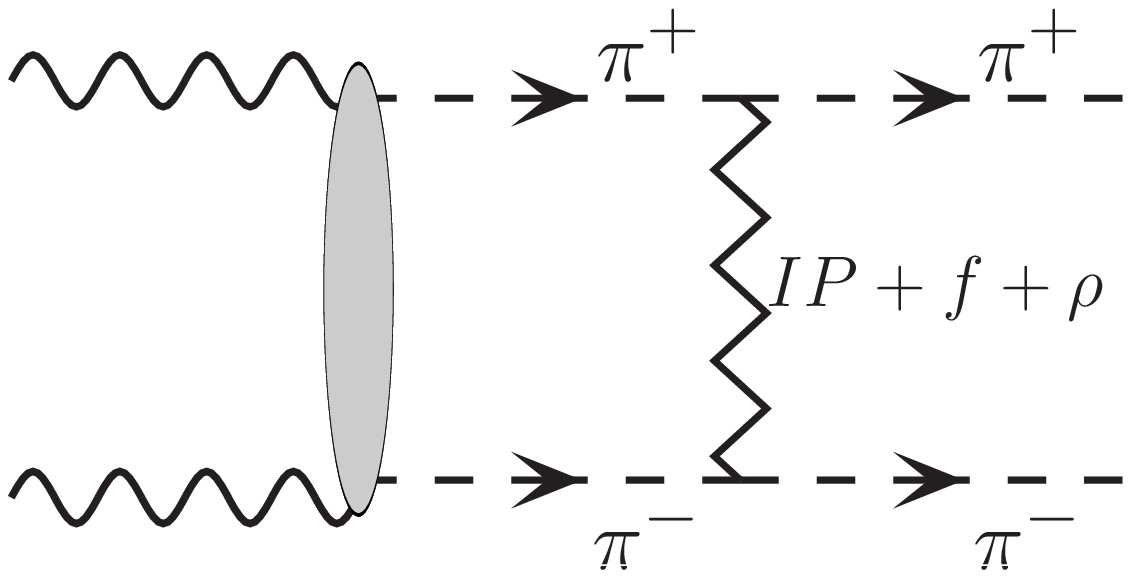}
  \includegraphics{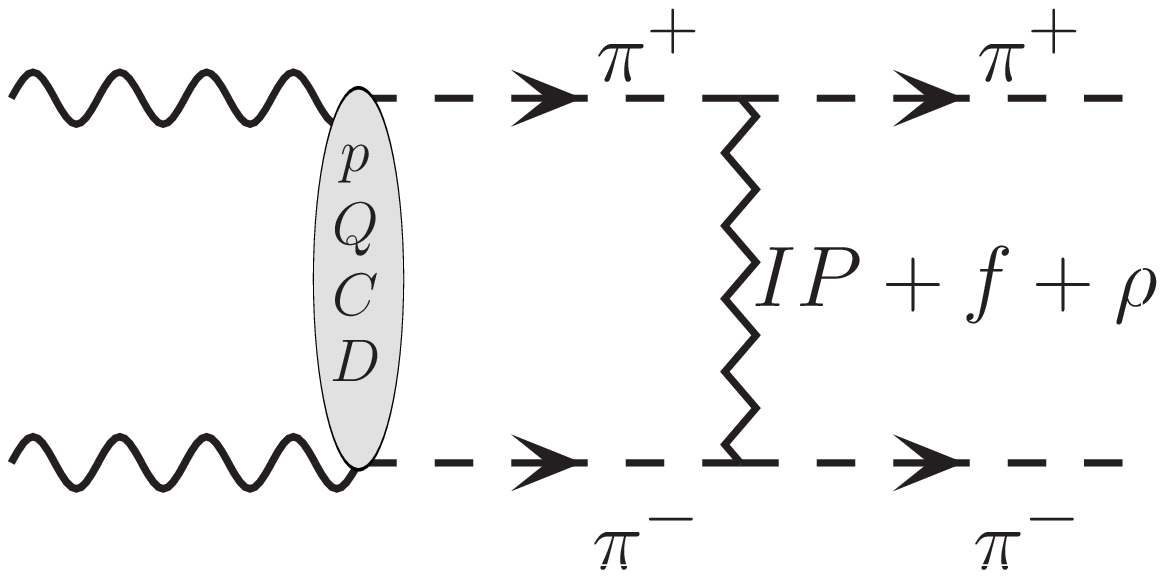} 
  \includegraphics{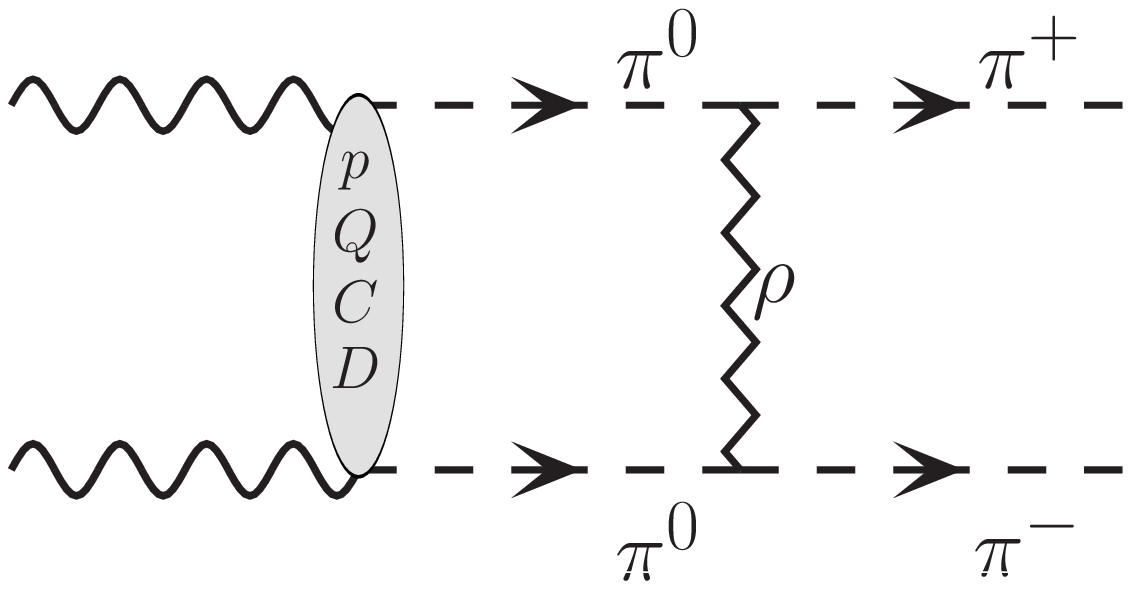}}
\caption{Pion-pion rescattering. }
\label{fig:el2}       
\end{figure}
\begin{figure}[h]
\centering
\resizebox{0.6\columnwidth}{!}{%
  \includegraphics{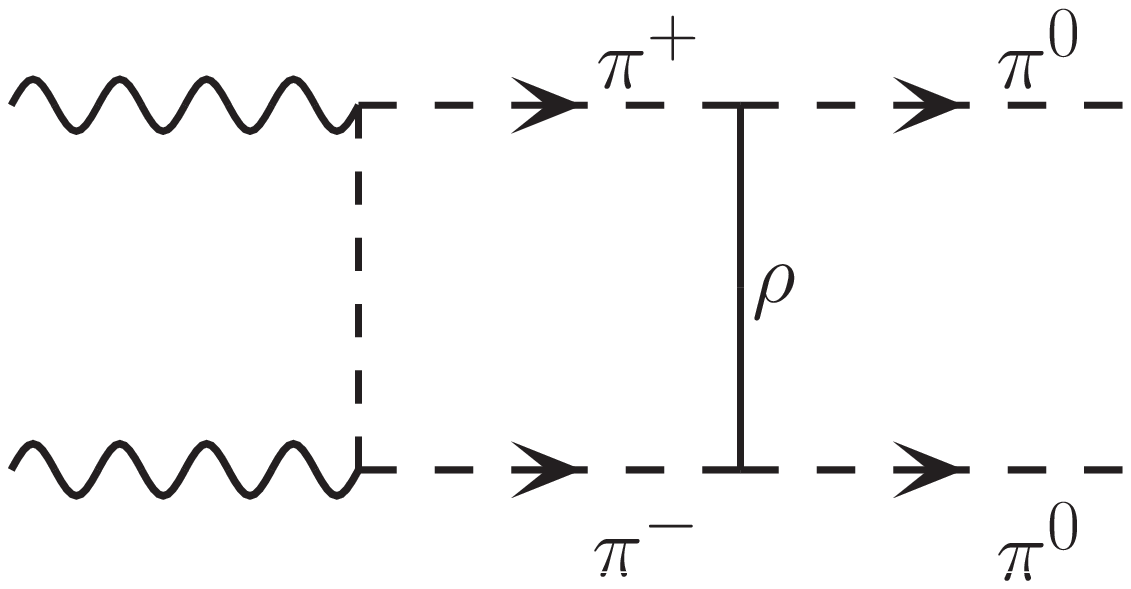}
\hspace{4cm}
  \includegraphics{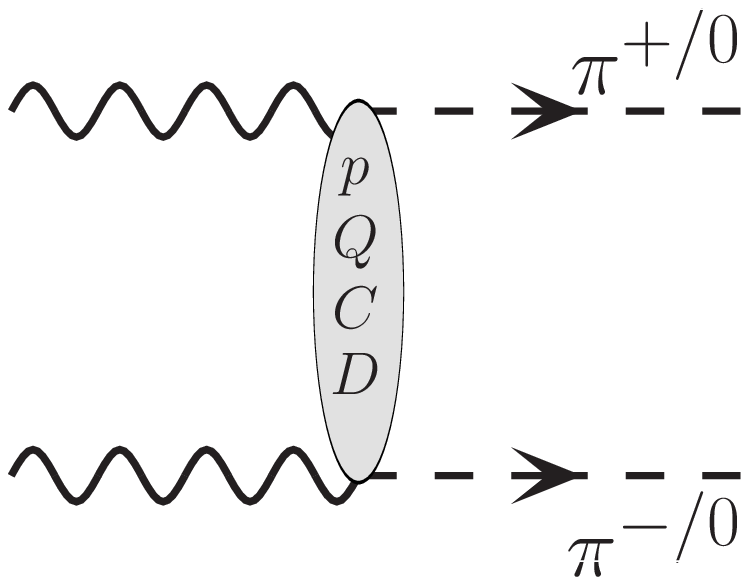} }
\caption{$\gamma\gamma\to\pi^0\pi^0$ 
in a simple coupled channel model with $\rho^\pm$ exchange (left panel)
and the Brodsky-Lepage perturbative mechanism 
for $\gamma\gamma\to\pi\pi$ scattering.}
\label{fig:el3}       
\end{figure}

Fig.~\ref{fig:elementary} shows 
the cross section function of $W$.
The angular ranges in the figure caption 
correspond to experimental cuts:
$|\cos\theta|<0.6$ for $\gamma\gamma\to\pi^+\pi^-$ and
$|\cos\theta|<0.8$ for neutral pion production. 
Several mechanisms are identified.
The dashed-dotted lines denote the contribution
from resonances (see Fig.~\ref{fig:el1}).
We take into account the following 
scalar: $f_0(600)$, $f_0(980)$,
tensor: $f_2(1270)$, $f'_2(1525)$, $f_2(1560)$ as well as 
one spin-4: $f_4(2050)$ resonances.
Our parametrization of the helicity dependent
resonant amplitude is a standart relativistic 
Breit-Wigner distribution with the energy-dependent
resonance widths using 
the spin dependent Blatt-Weisskopf form factor \cite{BW}.
\\
The solid blue line denotes 
$\gamma\gamma\to\pi^+\pi^-$ pion exchange Born continuum.
The helicity dependent amplitude is a sum of three terms 
(contact amplitude, $t$- and $u$-channel pion exchange amplitude)
multiplied by $t$ and $u$ dependent form factor.
Finally we use the QED amplitude for the real finite-size pions.
\\
The solid lines describe the high-energy pion-pion rescattering
(see Fig.~\ref{fig:el2}), $t/u$-channel $\rho^\pm$ meson exchange and 
pQCD Brodsky-Lepage mechanism (Fig.~\ref{fig:el3}).
The corresponding formalism of those mechanisms 
is rather complicated and
it will be presented in our forthcoming publication \cite{KGSz_future}.
The pQCD mechanism is very important 
for large invariant mass of the pion pairs.
The details related to this mechanism one can find 
in our last paper \cite{KS_pQCD}.
We can see that processes with the pomeron and reggeon exchange 
play a crucial role at low energy. 
In the same region of energy a major contribution 
to the final result gives $f_0(600)$ $s$-channel resonance
as well as $\rho$ meson exchange (pink solid line - right panel).
At arround  W $\approx 1$ GeV we observe a small dip.
This comes from interfering of different amplitudes.

\section{Nuclear cross section}
%
\begin{figure}[h]
\centering
\resizebox{0.3\columnwidth}{!}{%
  \includegraphics{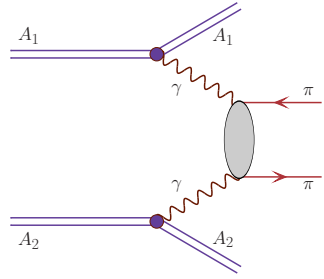} }
\caption{The Feynman diagram illustrating 
the formation of the pion pair. }
\label{fig:reaction}       
\end{figure}
In Fig.~\ref{fig:reaction} we show the basic mechanism 
of the exclusive electromagnetic pion pair production 
as a result of the peripheral nuclear collisions.
The equivalent photon approximation 
in the impact parameter space is the best suited approach
for applications to the peripheral collisions of nuclei.
This approach have been used recently in the calculation
of the 
$\rho^0 \rho^0$ pairs \cite{KS_rho}, 
$\mu^+ \mu^-$ \cite{KS_muon}, 
$Q \bar Q$ \cite{KS_quarks}, 
$D \bar D$ \cite{LS_DD} and for high mass 
dipions \cite{KS_pQCD}.
There one can find the details of the b-space EPA derivation.
Below we present a useful and compact formula for calculating 
the total cross section for $Pb Pb \to Pb \pi \pi Pb$ process:
  \begin{equation}
\sigma   \left(s_{NN} \right) =  
\int {\hat \sigma} \left(\gamma\gamma \rightarrow \pi \pi;W_{\gamma \gamma} \right) 
\theta \left(|{\bf b}_1-{\bf b}_2|-2R_A \right)
N \left( \omega_1, {\bf b}_1 \right ) N \left( \omega_2, {\bf b}_1 \right ) 
2 \pi b_m \, {\rm d} b_m \, {\rm d} \overline{b}_x \, {\rm d} \overline{b}_y 
\frac{W_{\gamma \gamma}}{2} {\rm d}W_{\gamma \gamma} {\rm d} Y \; . \nonumber
\label{eq.nuclear}
  \end{equation} 
This is a convolution 
of the elementary cross section and  photon fluxes.
In the EPA approach absorption effect is taken into account
by limiting impact parameter $b>R_1+R_2 \approx 14$ fm.
\begin{figure}[h]
\centering
\resizebox{0.85\columnwidth}{!}{%
  \includegraphics{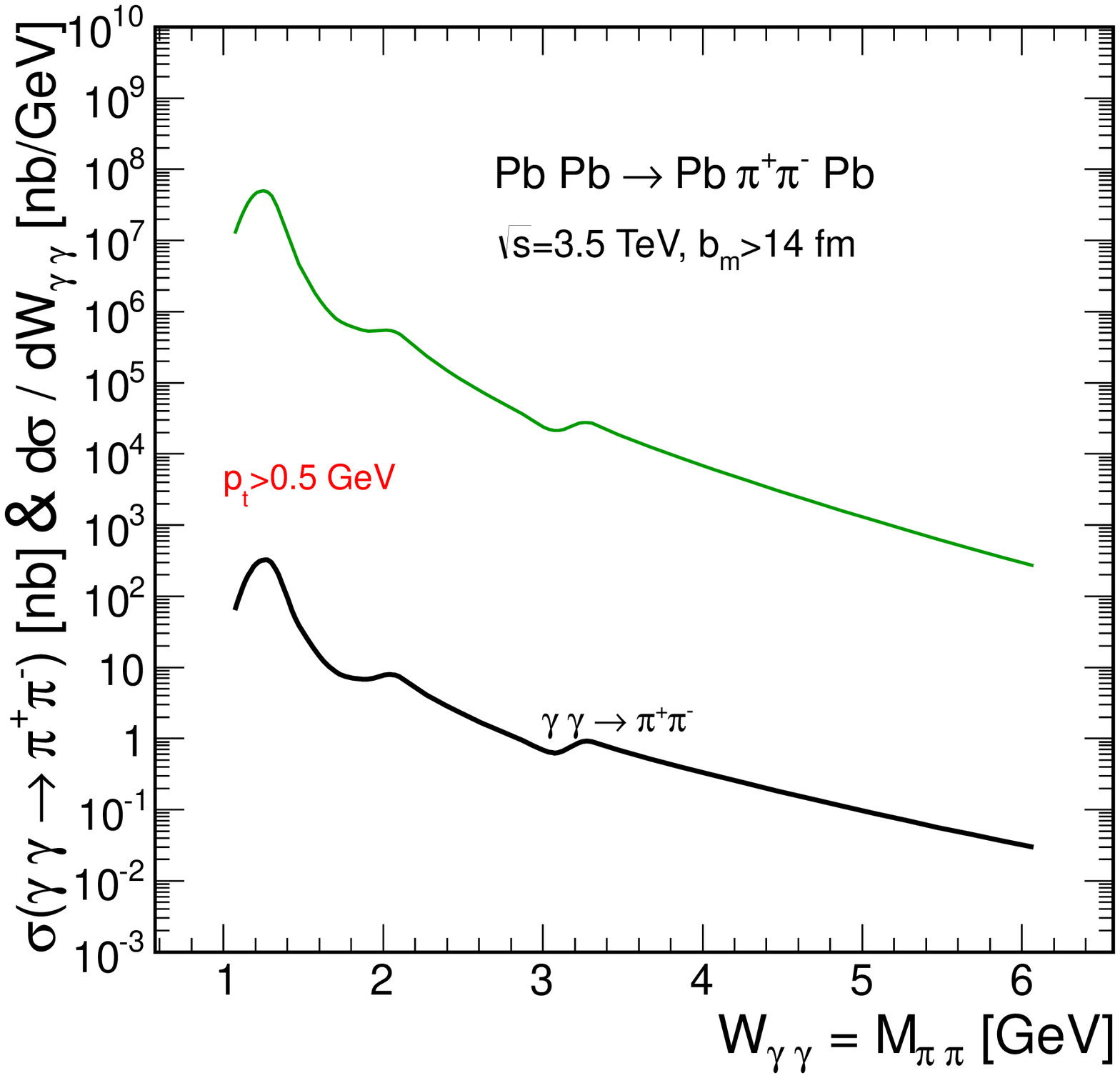}
  \includegraphics{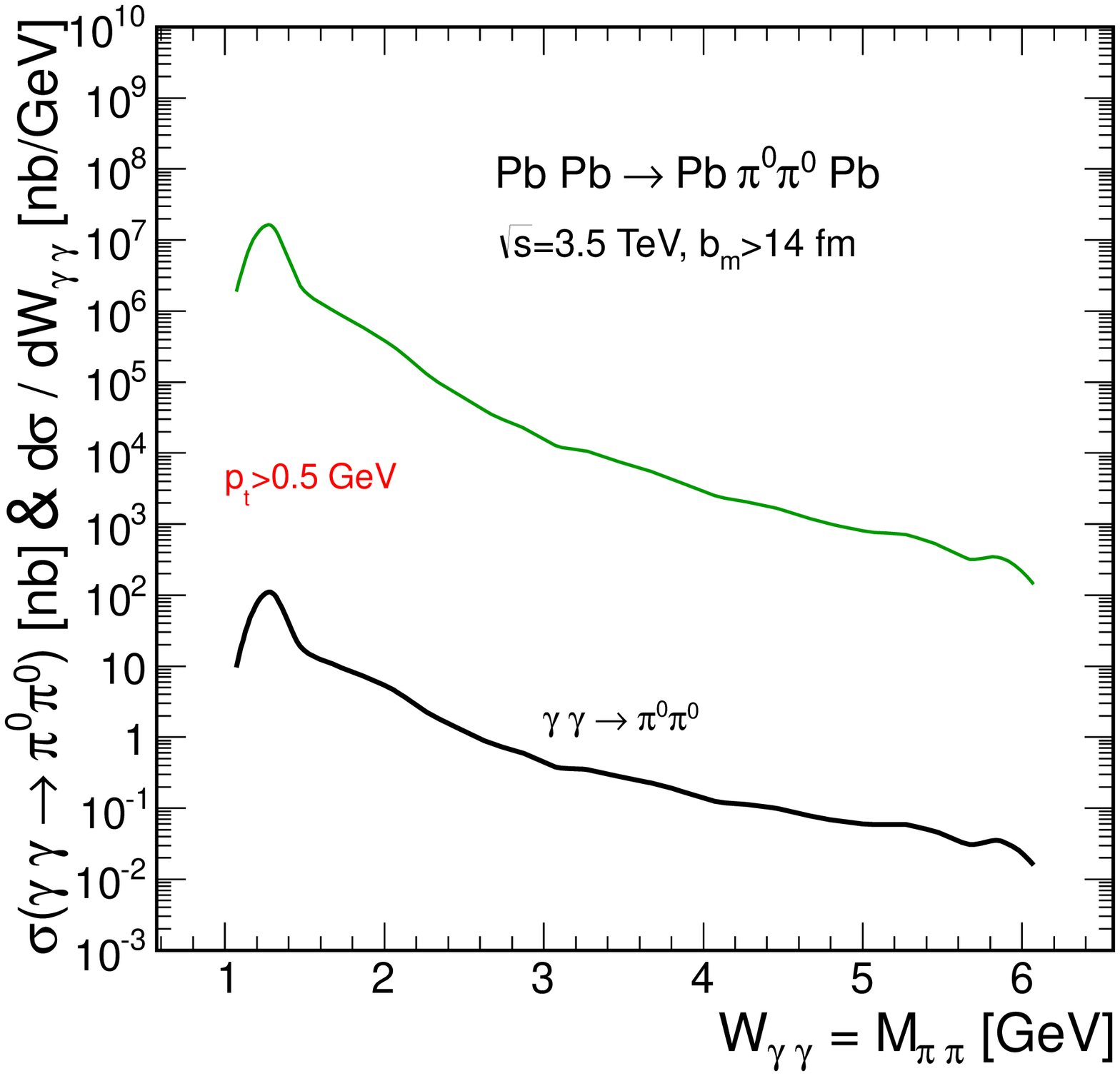} }
\caption{The nuclear cross section (green lines) 
as a function of $\gamma\gamma$ subsystem energy 
for the $Pb Pb \to Pb \pi^+\pi^- Pb$ (left panel) and
for the $Pb Pb \to Pb \pi^0\pi^0 Pb$ (right panel)
reactions calculated 
with an extra cut-off on pion transverse momentum.
The black lines refer to the elementary cross sections.}
\label{fig:dsig_dw}       
\end{figure}
\begin{figure}[h]
\centering
\resizebox{0.85\columnwidth}{!}{%
  \includegraphics{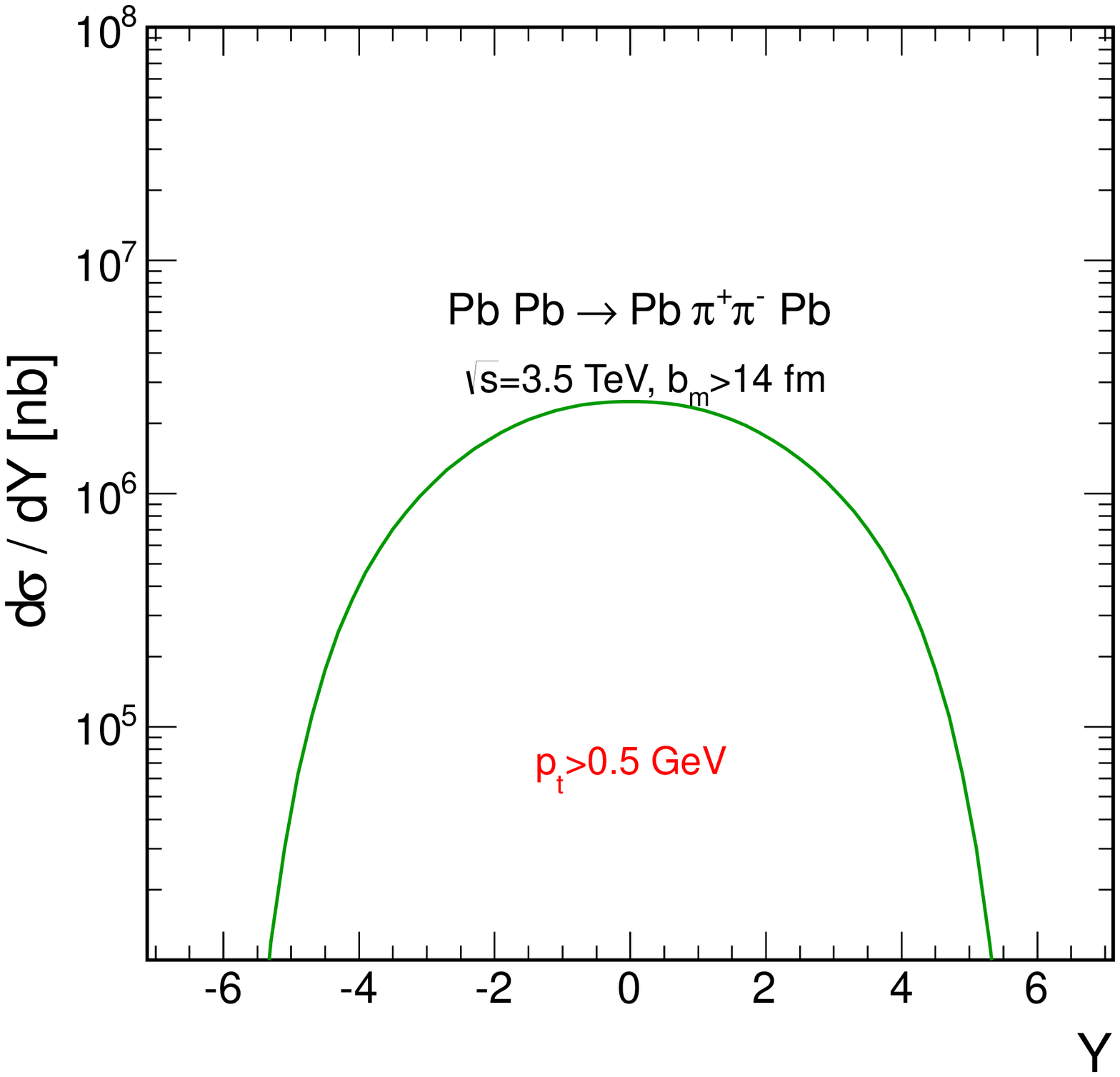}
  \includegraphics{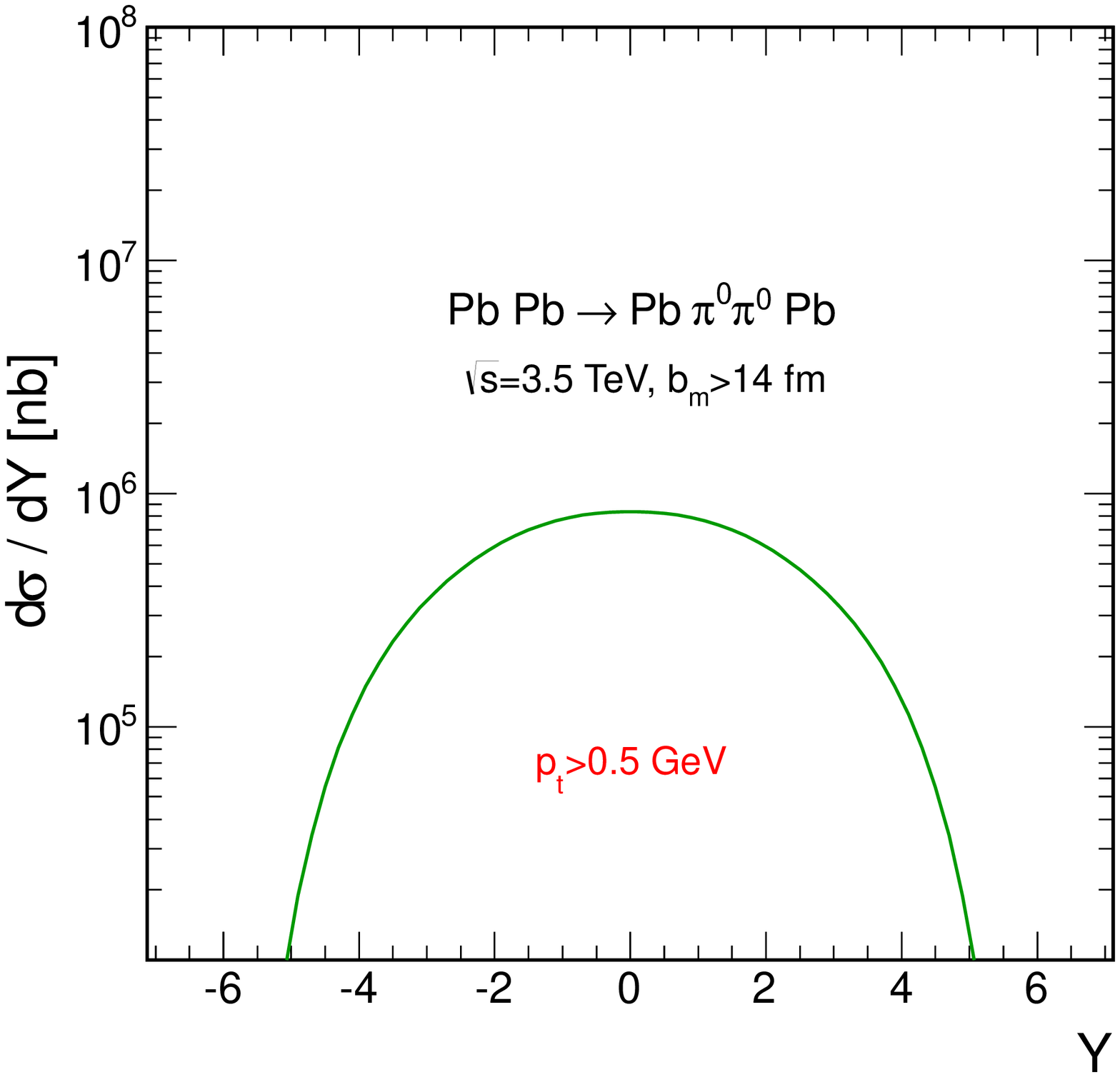} }
\caption{The nuclear cross section as a function 
of pion pair rapidity
for the $Pb Pb \to Pb \pi^+\pi^- Pb$ (left panel) and
for the $Pb Pb \to Pb \pi^0\pi^0 Pb$ (right panel)
reactions calculated 
with an extra cut-off on pion transverse momentum.}
\label{fig:dsig_dy}       
\end{figure}

In Fig.~\ref{fig:dsig_dw}, \ref{fig:dsig_dy} we present 
the predictions for possible future experiments.
We show total (angle-integrated) cross section 
as a function of $\gamma\gamma$ energy 
as well as as a function of pion pair rapidity
for both charged and neutral pions.
We consider lead-lead collisions 
at energy $\sqrt{s_{NN}}=3.5$ TeV.
In these calculations we have imposed extra cuts 
on pion transverse momenta (p$_t>0.5$ GeV).
The nuclear cross section is a reflection of 
the elementary cross section. 
We have obtained measurable cross sections. 

\section{Conclusion}
%
We have discussed a possibility to study 
the $\gamma\gamma\to\pi\pi$ processes in ultraperipheral
ultrarelativistic heavy-ion collisions. 
First, we have described the experimental data
for the $\pi\pi$ production in $e^+e^-$ collisions.
To obtain correct description of the data
we have taken into account pion exchange 
for $\pi^+\pi^-$ channel, several resonances, 
pQCD mechanisms, pion-pion rescatterings 
both at low and intermediate energies. 
Next, cross sections for corresponding nuclear
collisions was calculated.
The cross sections are fairly large.
Whether they can be measured 
requires further studies.
\begin{table}[h]
\caption{The total cross section 
for the $Pb Pb \to Pb \pi\pi Pb$ reactions.}
\label{tab:1}       
\centering
\begin{tabular}{lcr}
\hline\noalign{\smallskip}
Reaction & Extra cut on $p_t$ & Total cross section  \\
\noalign{\smallskip}\hline\noalign{\smallskip}
$Pb Pb \to Pb Pb \pi^+ \pi^-$ &  & 112 mb  \\
$Pb Pb \to Pb Pb \pi^+ \pi^-$ & $p_t>0.5$ GeV & 14 mb \\
$Pb Pb \to Pb Pb \pi^0 \pi^0$ &  & 10 mb \\
$Pb Pb \to Pb Pb \pi^0 \pi^0$ & $p_t>0.5$ GeV & 5mb \\
\noalign{\smallskip}\hline
\end{tabular}
\end{table}
\\
This work was partially supported by the Polish grant N N202 236640.


\begin{thebibliography}{}

\bibitem{KS_rho}
M. K{\l}usek, A. Szczurek, and W. Sch\"afer,
Phys. Lett. {\bf B674}, (2009) 92.

\bibitem{KS_muon}
M. K{\l}usek-Gawenda and A. Szczurek,
Phys. Rev. {\bf C82}, (2010) 014904.

\bibitem{KS_quarks}
M. K{\l}usek-Gawenda, A. Szczurek, M. V. T. Machado and V. G. Serbo,
Phys. Rev. {\bf C83}, (2011) 024903.

\bibitem{LS_DD}
M. {\L}uszczak and A. Szczurek,
Phys. Lett. {\bf B700}, (2011) 116.

\bibitem{KS_pQCD}
M. K{\l}usek-Gawenda and A. Szczurek,
Phys. Lett. {\bf B700} (2011) 322.

\bibitem{BW}
J. M. Blatt and V. Weisskopf,
\textit{Theoretical Nuclear Physics}
(John Wiley, New York 1952).

\bibitem{KGSz_future}
M. K{\l}usek-Gawenda, A. Szczurek, a paper in preparation.
\end{thebibliography}
\end{document}